\documentclass[superscriptaddress]{revtex4}
\usepackage{graphicx}
\usepackage{amsmath}
\usepackage{amssymb}
\usepackage{slashed}
\usepackage{latexsym}
\usepackage{epsfig}
\usepackage{amsbsy}
\usepackage{array}
\usepackage{amssymb}
\usepackage{changes}
\usepackage{color}
\usepackage{setspace}
\usepackage{bm}
\usepackage{lipsum}
\usepackage{mathrsfs}
\usepackage{float}
\usepackage{color}
\usepackage[T1]{fontenc}
\usepackage{mathptmx}
\usepackage{caption2}
\usepackage{subfigure}
\usepackage{hyperref}

\begin{document}
	
	\title{ Exploring the parameter space of quasi-particle model with the strange quark stars}
	
	\author{Wen-Hua Cai }
	\email{stuwenhua.cai@gmail.com }
	\affiliation{College of physics, Sichuan University, Chengdu 610064, China}
	
	\author{Qing-Wu Wang}
	\email{ qw.wang@scu.edu.cn}
	\affiliation{College of physics, Sichuan University, Chengdu 610064, China}

\begin{abstract}
The properties of strange quark stars are studied within the   quasi-particle model. Taking into account the chemical equilibrium and charge neutrality, the EOS of $ (2+1) $-flavor quark matter is obtained.  We illustrate the parameter spaces with constraints from two aspects: the one is based on the astronomical results of PSR J$ 0740+6620 $ and GW$ 170817 $, and another is based on the constraints proposed from the theoretical study of compact star that the EOS must ensure the tidal deformability $ \varLambda_{1.4}=190^{+390}_{-120} $ and support a maximum mass above $ 1.97~M_{\odot} $. It is found that the both types of constraints can not restrict the parameter space of quasi-particle model in a reliable region and thus we conclude that the small compact star cannot be strange quark star.


\end{abstract}
 \maketitle
\section{Introduction}
Quantum Chromodynamics (QCD), which has been regarded as  the theory to govern the strong interaction, predicts that the confined hadronic matter will undergo a deconfinement transition to a new phase at large density and extremely low temperature \cite{1,2}.
It is believed that the new phase of nuclear matter, as we call it quark-gluon plasma (QGP), may exist within a new kind of compact stars--quark stars \cite{3,4,5,6,7,8}. The strange quark matter (SQM) hypothesis, indicates that the matter formed by deconfined $u$ quark, $d$ quark and $s$ quark may be the true ground state of strongly interacting matter \cite{Terazawa,24}. This hypothesis support the presence of strange quark star.

 In the recent years, basing on the latest astronomical data, some researchers argued that many compact stars and pulsars with large mass may be strange quark stars, even including some stars which used to be identified as neutron stars \cite{55}. So it naturally  raises the question how   to discriminate these strange quark stars from other compact stars.  At present, since the lack of astronomically observation data, it  may possibly only through analysis of properties, like the mass-radius relation and  tidal deformability, which heavily depends on the studying of the equation of state (EOS)\cite{9,10}.
Applying an EOS into the Tolman-Oppenheimer-Volkoff (TOV) equation, the relevant properties and features of the star can be obtained. However at high density and very low temperature, it is difficult to figure out the EOS of cold deconfined quark matter from the first principles. Hence, people study the properties of QGP and strange quark stars basing on the nonpertrubative   models like the MIT bag model \cite{11,12,13,14}, the Nambu$-$Jona-Lasinio (NJL) model \cite{15,16,17, 19,20}, the quasi-particle model \cite{20,21,22} and etc.

The quasi-particle model is a kind of phenomenological description of QGP  which were obtained by the simulation of the Lattice QCD \cite{23}. Via the introduction of thermal mass at finite temperature and finite chemical potential, we can treat the system as  made up of free quark gas to avoid the complicated  calculation of  QCD interactions.
Thanks to the increasing data from astrophysical observations, it provides a place to test the   applicability of effective  model.   The latest astrophysical observation of pulsar, PSR J$0740+6620$ $(M=2.14^{+0.10}_{-0.09} ~M_{\odot})$ \cite{25}, has provided a maximum mass measurement result so far, which is much larger than the result of PSR J$0348+0432$ $(M=2.01\pm 0.04 ~M_{\odot})$ \cite{26}, and thus the soft EOSs which can not produce such a massive star are supposed to be eliminated.
Besides, according to the gravitation wave observation GW$ 170817$ \cite{27}, the tidal deformability $ \varLambda $ for 1.4 $ M_{\odot} $ star $ (\varLambda_{1.4}) $ needs to be constricted to smaller than 800 in the low-spin prior case. Recently, some significant conclusions have been drawn via the study on the quark matter within the compact stars taking into account both the astronomical observations and theoretical simulations \cite{28}. Adopting the speed-of-sound interpolation method, they  have succeeded in describing the QCD matter properties of different compact stars with different masses and radii. And their results have been verified by approximately $ 570000 $ EoSs which are built from randomly generated functions. To get such a  result,   some hypothetical constraints are proposed that the maximum mass should be above $ 1.97 $ $ M_{\odot} $ and the range of tidal deformability should be revised to $ \varLambda_{1.4}=190^{+390}_{-120} $. Lately, some scholars even claim that there may exists compact star with a mass of $ 2.6 $ $ M_{\odot} $ \cite{29,30}, which presents a great potential to be the candidate of strange quark star if it is proved to be true. In this paper, in the viewpoint of strange quark stars, we intend to study the parameter space of quasi-particle model with some latest astrophysical results. 

This paper is organized as follows.   We first introduce the quasi-particle model at finite chemical potential basing on the statistical mechanics and thermodynamic equilibrium in section \ref{sec2}. By considering the chemical equilibrium and charge neutrality, the EOS of SQM can be obtained in   section \ref{sec3}. Then, making use of the the EOS, the mass-radius and tidal deformability  of strange quark star are explored . On the basis of the latest astronomical observations and theoretical researches, we study the parameter space of quasi-particle model. At last, a short summary of our work is given in  section \ref{sum}.

\section{The quasi-particle model at finite chemical potential and zero temperature}
\label{sec2}
The quasi-particle model, which was proposed to explain the   results in lattice gauge theory(LGT) simulations at first, is widely used to describe the non-perturbative behavior of QCD. To illustrate the quark matter of $(2+1)$-flavor QGP, we construct the quasi-particle model basing on statistical mechanics and thermodynamic equilibrium. Here, we start from the condition of finite chemical potential $ \mu $ and finite temperature $T$. In this case, the density of quarks is given by
\begin{equation}\label{Eq.1}
\rho_{i}(T,\mu)=2N_{c}\int\frac{\mathrm{d}^{3}k}{(2\pi)^{3}}\left(\frac{1}{e^{\frac{(\omega_{i}-\mu)}{T}}+1}+\dfrac{1}{e^{\frac{(\omega_{i}+\mu)}{T}}+1}\right) ,
\end{equation}
where $ N_{c} $ is the number of colors and subscript $i$ indicates the $u$, $d$ and $s$ quarks. In addition,
\begin{equation}\label{Eq.2}
\omega_{i}(T,\mu)=\sqrt{k^{2}+m^{2}_{i}(T,\mu)}
\end{equation}
is the dispersion relation for each kind of quarks, where $ m_{i}(T,\mu) $ stands for the effective mass of different quarks. Following the works of Bannur \cite{31,32,33,34,35}, $ m_{i}(T,\mu) $ have the expression of
\begin{equation}\label{Eq.3}
m_{i}^{2}(T,\mu)=(m_{i0}+m_{th}(T,\mu))^{2}+m_{th}^{2}(T,\mu),
\end{equation}
in which $ m_{i0} $ represents the current quark mass  and $ m_{th}(T,\mu) $ is the thermal mass term that stands for the complicated QCD interaction. In this paper, we take $ m_{s0}=150$ MeV  and $ m_{u0, d0}=m_{s0}/28.15\approx5.33$ MeV
in the light of   \cite{36}.
Then, taking the limit $ T\to0 $ and computing the integral of  \eqref{Eq.1} from zero to Fermi momentum $ k_{F}=\sqrt{\mu^{2}-m_{i}(\mu)^{2}} $, the number density of quarks
\begin{equation}\label{Eq.6}
\rho_{i}(\mu)=\frac{N_{f}}{3\pi^{2}}(\mu^{2}-m^{2}_{i}(\mu))^{3/2}\theta(\mu-m_{i}(\mu))
\end{equation}
can obtained. The symbol $ \theta $ in   \eqref{Eq.6} stands for the step function.

In this case, the thermal mass $ m_{th} $ in   \eqref{Eq.3} reads
\begin{equation}\label{Eq.7}
m_{th}^{2}(\mu)=\frac{N_{f}\mu^{2}g^{2}(\mu)}{18\pi^{2}},
\end{equation}
in which the symbol $ N_{f} $ is number of flavors.
The effective coupling constant $ g $ can be obtained by the two-loop approximation and has the form of
\begin{equation}\label{Eq.8}
\begin{aligned}
g^{2}(\mu)&=4\pi\alpha_{s}(\mu)=\frac{24\pi^{2}}{(33-2N_{f})\ln(1.91\mu/2.91\zeta)}\\
&\times\left[1-\frac{3(153-19N_{f})}{(33-2N_{f})^{2}}\frac{\ln(2\ln(1.91\mu/2.91\zeta))}{\ln(1.91\mu/2.91\zeta)}\right],
\end{aligned}
\end{equation}
 where $\zeta$ is a phenomenological parameter which is related to the non-perturbative effect of QCD \cite{59,60}. Making use of the physical quantities introduced above and the basic thermodynamic relations, all the physical quantities we need to describe the SQM can be derived.

\begin{figure}[htbp]
   \centering
	\includegraphics[width=0.85\textwidth]{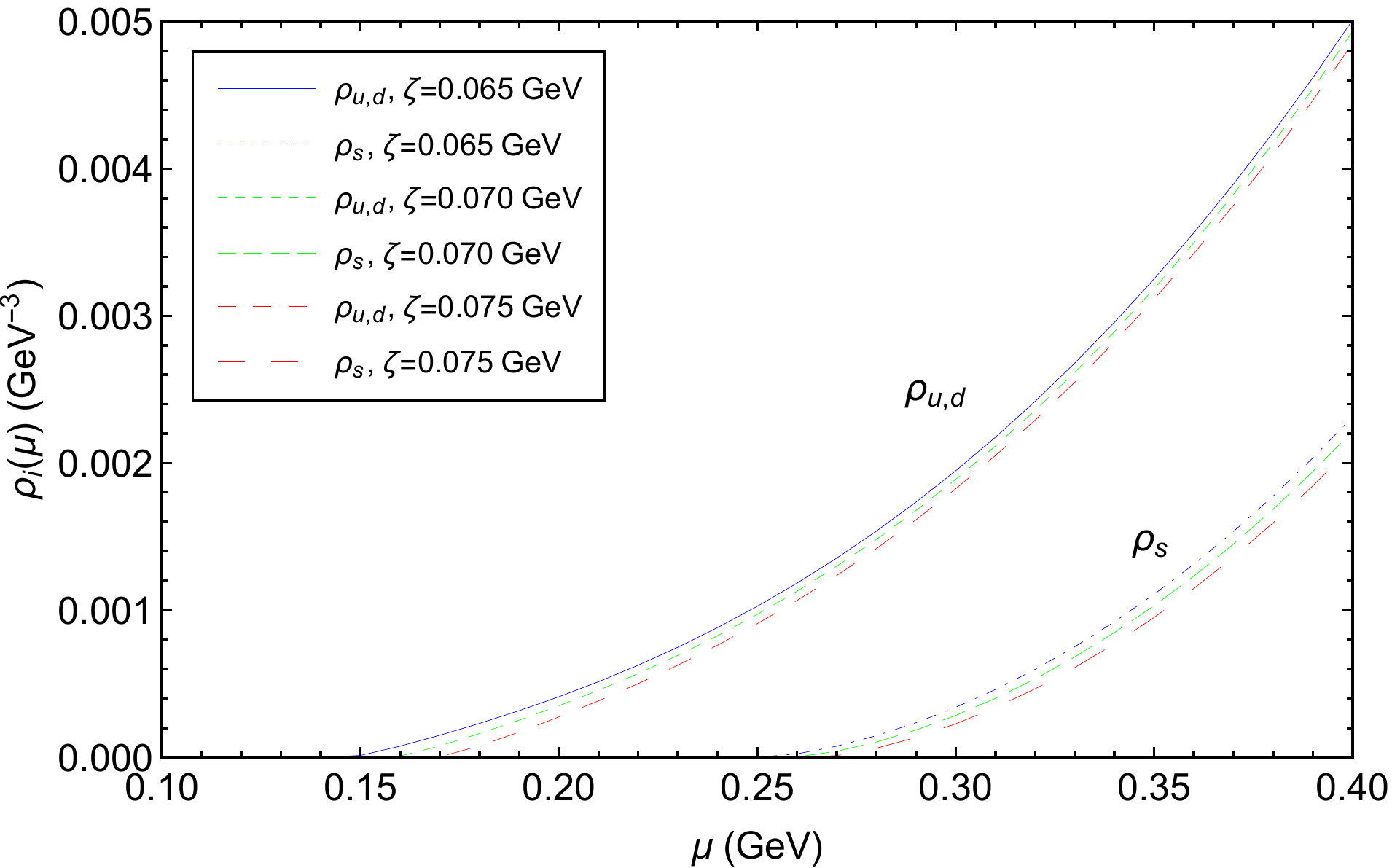}
	\caption{The number density $\rho_{i}(\mu)$ of $u$, $d$ and $s$ quark as functions of chemical potential $ \mu $ at zero temperature with $\zeta=0.065 $ GeV, $0.070$ GeV and $0.075$ GeV respectively. }
	\label{fig.1}
\end{figure}

Solving   \eqref{Eq.3}, \eqref{Eq.6}, \eqref{Eq.7} and \eqref{Eq.8}, the relation between number density of quark $\rho_{i}(\mu)$ and chemical potential $ \mu $ can be obtained, as showed in figure \ref{fig.1}. Owing to the step function in the expression of $\rho_{i}(\mu)$, it is found that the quark number density vanishes when the chemical potential is below a critical point $ \mu_{c} $ in this figure. That is to say, $ \mu=\mu_{c} $ is a singularity which divides the quark number density into two different regions. This phenomenon is in agreement with the conclusion brought forth in   \cite{37}, in which researchers pointed out the existence of some singularity at critical point with zero temperature is a robust and model-independent result basing on a universal argument. Similar discussions have also been taken in   \cite{38,39}.

\section{The structure of strange quark stars with quasi-particle model }
\label{sec3}
At present, the research of strange quark stars has attracted increasing attention of scientists. The existing theories suggest the strange quark stars have at least two main channels to be produced. First, a number of them may come into being in the early stage of universe according to the big bang theory. Second, they also can be produced by the phase transition within the neutron stars \cite{55}. Therefore, studying the strange quark stars does  help to understand the phase transition from confined hadron matter to quark matter.

\begin{figure}[htbp]
	\centering
	\includegraphics[width=0.85\textwidth]{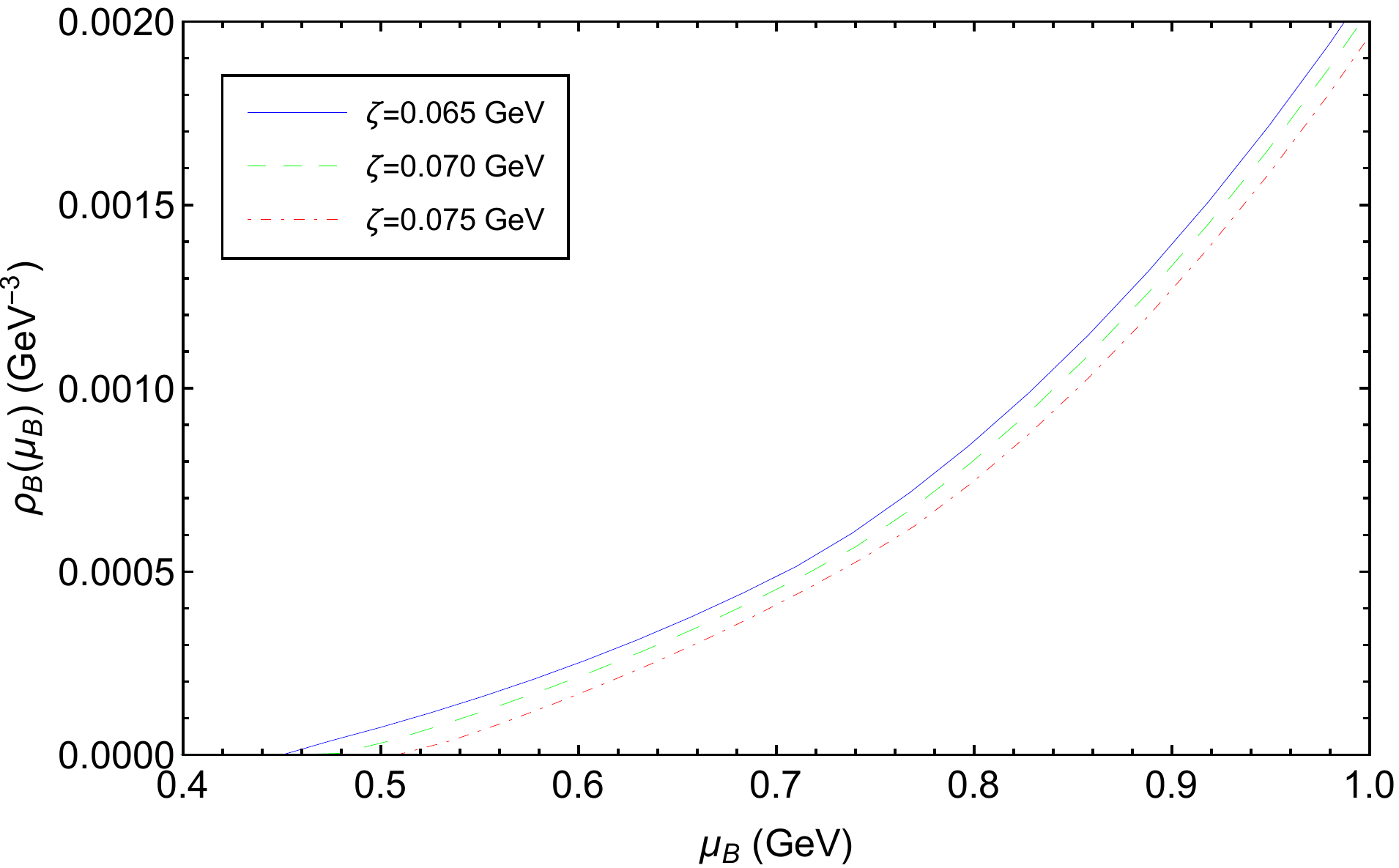}
	\caption{The baryon number density $ \rho_{B}(\mu_{B}) $ as a function of the baryon chemical potential $ \mu_{B} $ with  $ \zeta=0.065 $ GeV, $ 0.070 $ GeV and $ 0.075 $ GeV, respectively.}
	\label{fig.2}
\end{figure}

\begin{figure}[htbp]
\centering
\includegraphics[width=0.85\textwidth]{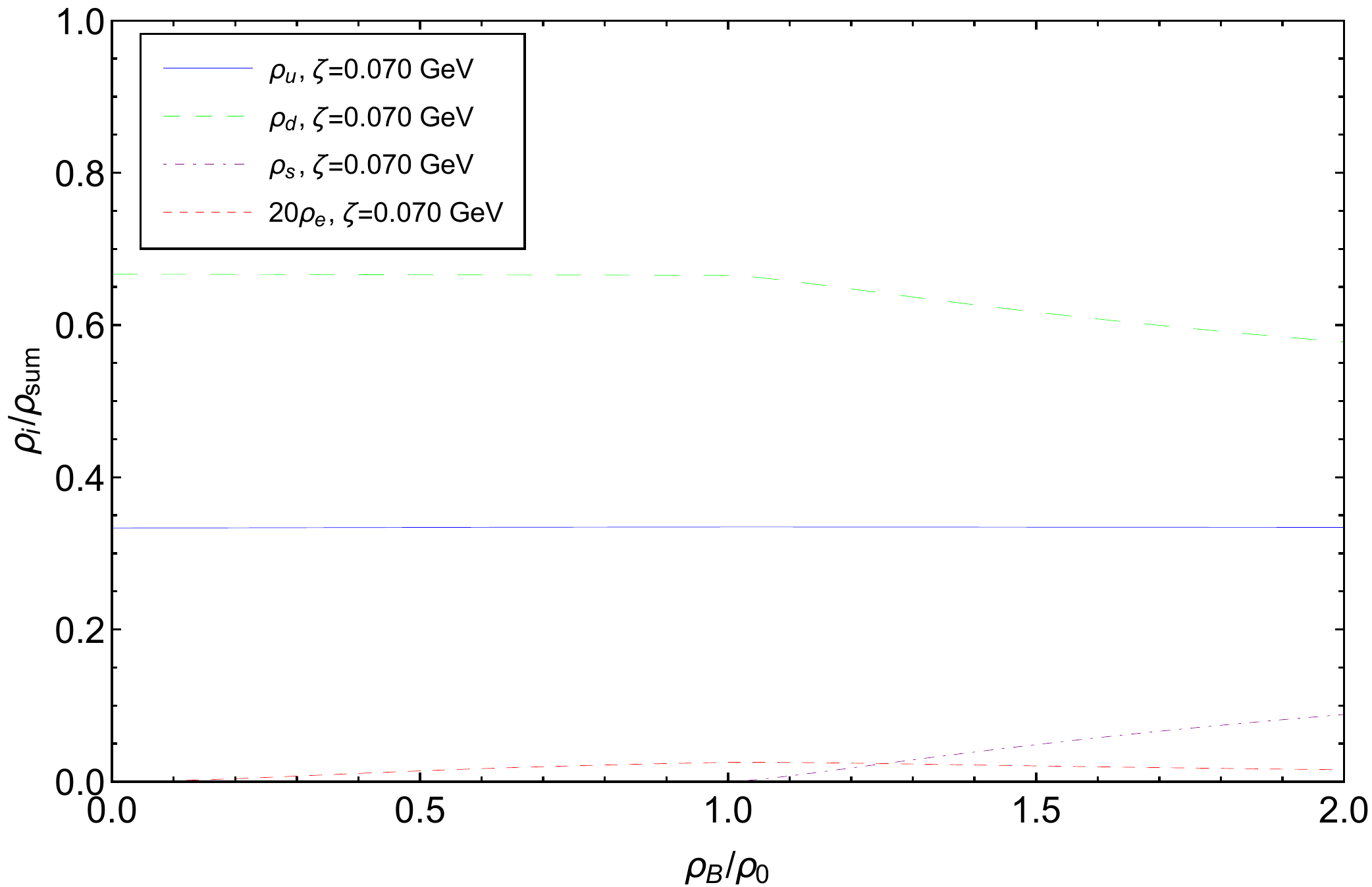}
\caption{The proportion of constituents $ u $, $ d $, $ s $ quarks and electrons (multiplied a factor of $ 20 $) $ \rho_{i}/\rho_{sum} $ as functions of the baryon number density $ \rho_{B}/\rho_{0} $ (the symbol $ \rho_{sum} $ stands  for the total quark number density and $ \rho_{0} $ stands for the baryon number density where the $ s $ quarks begin to be nonzero). }	
\label{fig.3}	
	
\end{figure}

To begin with, there exists chemical equilibrium with the compact stars  which is established via weak decay
\begin{equation}\label{Eq.9}
d\leftrightarrow u+e^{-}+\bar{\nu}_{e}\leftrightarrow s.
\end{equation}
In consideration of the equilibrium, the electrons and neutrinos have to be included in the system. However, for a stable quark star, we can reasonably assume the neutrinos have enough time to leave the system, which means that there is no need to consider  the participation of neutrinos in the chemical equilibrium. To that end, the constraint conditions
\begin{equation}\label{Eq.10.1}
\mu_{d}=\mu_{u}+\mu_{e}
\end{equation}
and
\begin{equation}\label{Eq.10.2}
\mu_{s}=\mu_{u}+\mu_{e}
\end{equation}
for the chemical potential of $u$, $d$, $s$ quarks and electrons have to be met. Moreover, taking the electric charge neutrality into account, the number densities of quarks and electron should be bound to satisfy the relation of
\begin{equation}\label{Eq.11}
\frac{2}{3}\rho_{u}-(\frac{1}{3}\rho_{s}+\frac{1}{3}\rho_{d}+\rho_{e})=0,
\end{equation}
where the electron density $ \rho_{e} $ is given by $ \rho_{e}=\mu^{3}_{e}/(3\pi^{2}) $ at zero temperature. Then, the relation between baryon number density $ \rho_{B}=(\rho_{u}+\rho_{d}+\rho_{s})/3 $ and baryon chemical potential $ \mu_{B}=\mu_{u}+\mu_{d}+\mu_{s} $, as well as the variation of constituents, can be obtained, which are shown in figure \ref{fig.2} and figure \ref{fig.3} respectively.
In figure \ref{fig.2}, we can see clearly that the baryon number density $ \rho_{B}(\mu_{B}) $ stays zero when the the baryon chemical potential $ \mu_{B} $ smaller than a critical point $ \mu_{Bc} $; while in the region of $ \mu_{B}>\mu_{Bc} $, $ \rho_{B}(\mu_{B}) $ becomes a monotonically increasing function of $ \mu_{B} $. 
And the proportion of constituents, including $u$, $d$, $s$ quarks and electrons, as functions of total baryon number density $ \rho_{B} $ in weak decays has been shown in figure \ref{fig.3} (the symbol $ \rho_{sum} $ stands for the total quark number density and symbol $ \rho_{0} $ stands for the baryon number density where the $ s $ quarks begin to be nonzero).
 We can see that the graph which represents the proportion of electrons $ \rho_{e}/\rho_{sum} $ just can be visible after the fraction multiplied a factor of $ 20 $. It means that $ \rho_{e}/\rho_{sum} $ stays in an extremely small range in SQM system.
Meanwhile, the proportion of $ u $ quarks is fixed to about $1/3$ as a consequence of charge neutrality. Thus, the $ d $ quarks occupy the remaining $2/3$ before the appearance of $s$ quarks. This consequence agrees with the conclusion in   \cite{1}, in which the researchers studied the same topic with NJL model. By the way, we find that the variation of constituents is not associated with the $ \zeta $. Therefore, we think such a behavior of constituents is a model-independent result and only depends on the chemical equilibrium and charge neutrality.

\begin{figure}[htbp]
	\centering
	\includegraphics[width=0.85\textwidth]{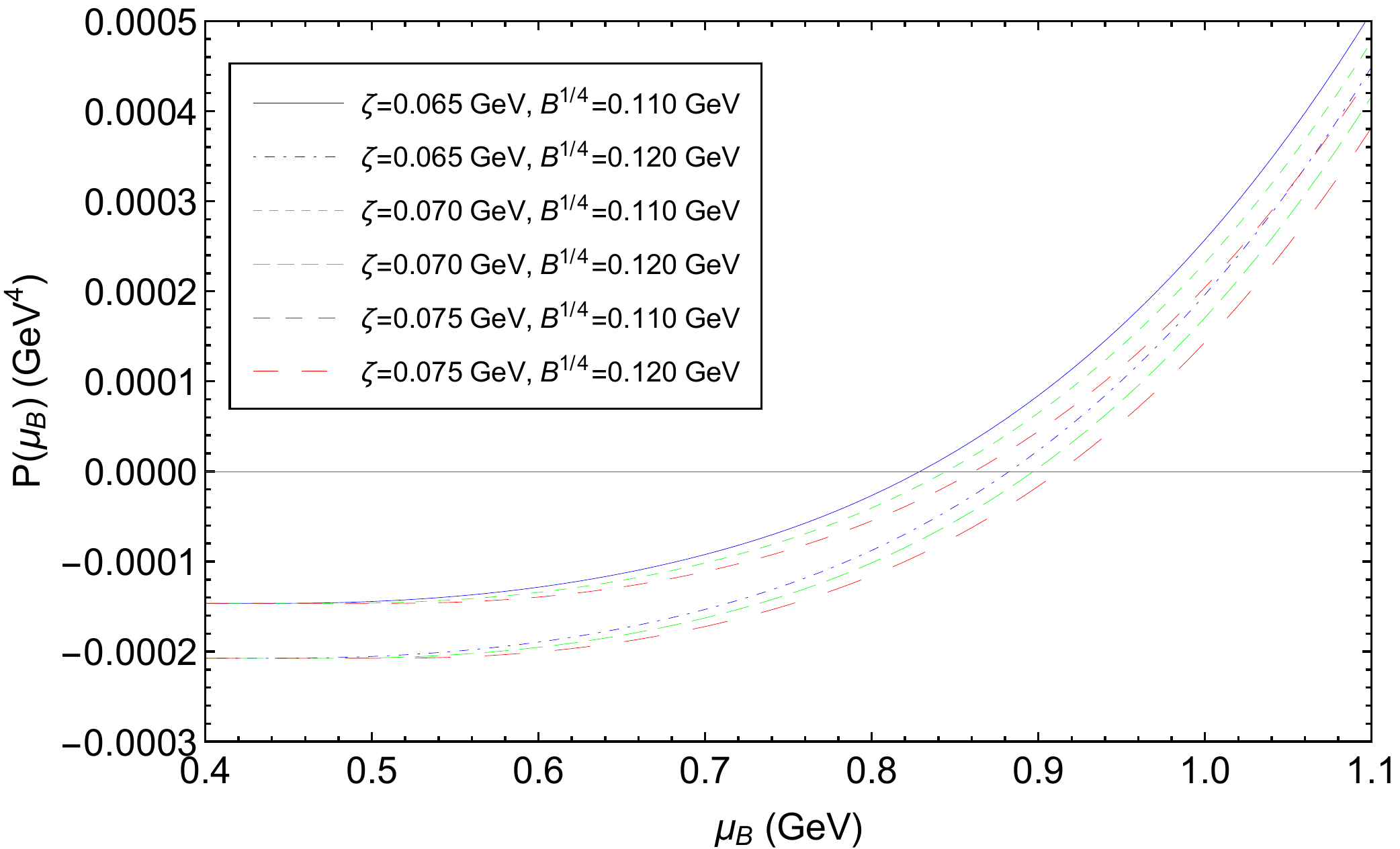}
	\caption{The pressure density $ P(\mu_{B}) $ as a function of baryon number density $ \mu_{B} $ for different $ \zeta $ and $ B $.}
	\label{fig4}
\end{figure}

\begin{figure}[htbp]
	\centering
	\includegraphics[width=0.85\textwidth]{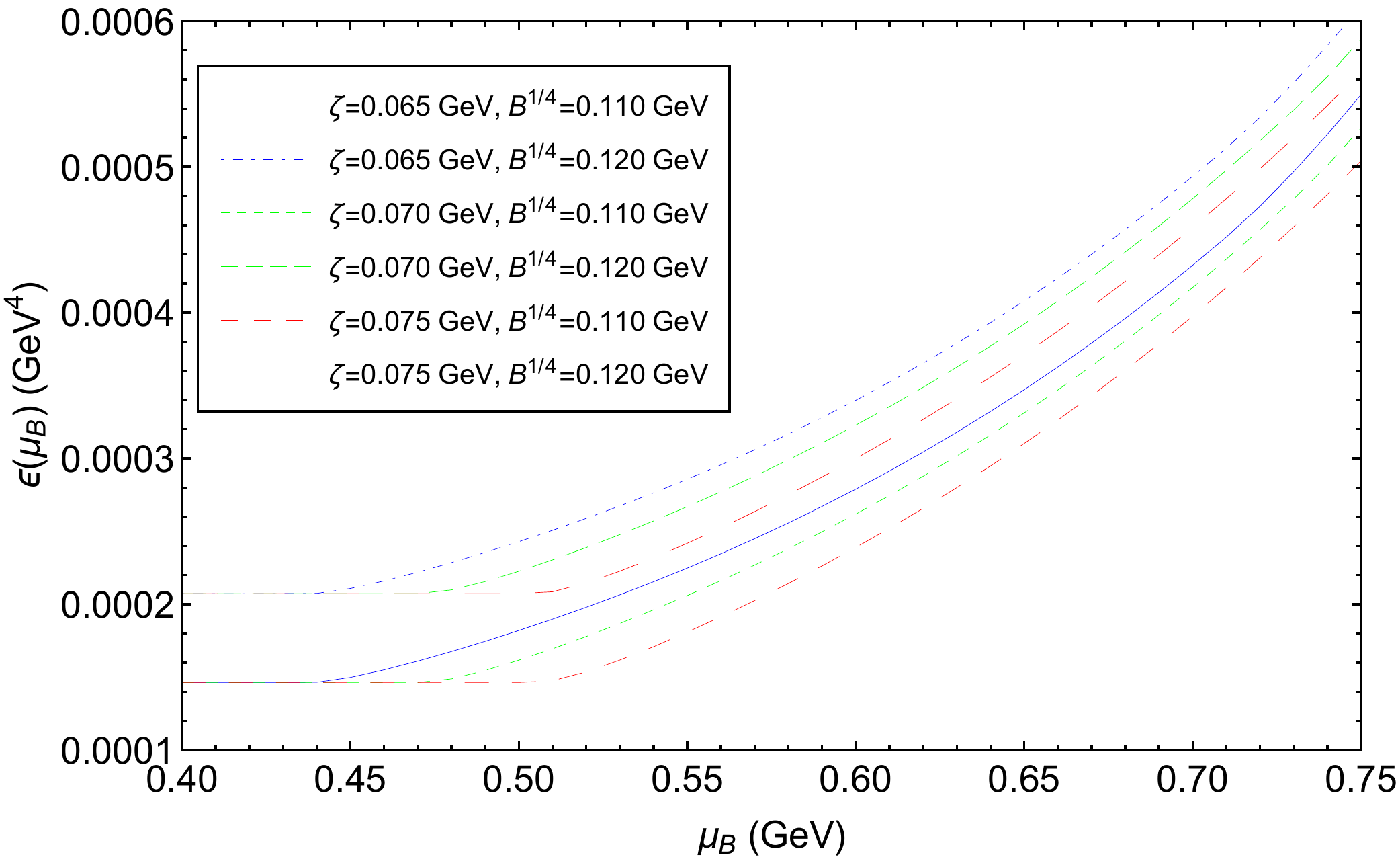}
	\caption{The energy density $ \varepsilon(\mu_{B}) $ as a function of baryon number density $ \mu_{B} $ for different $ \zeta $ and $ B $. }
	\label{fig5}
\end{figure}

\begin{figure}[htbp]
	\centering
	\includegraphics[width=0.85\textwidth]{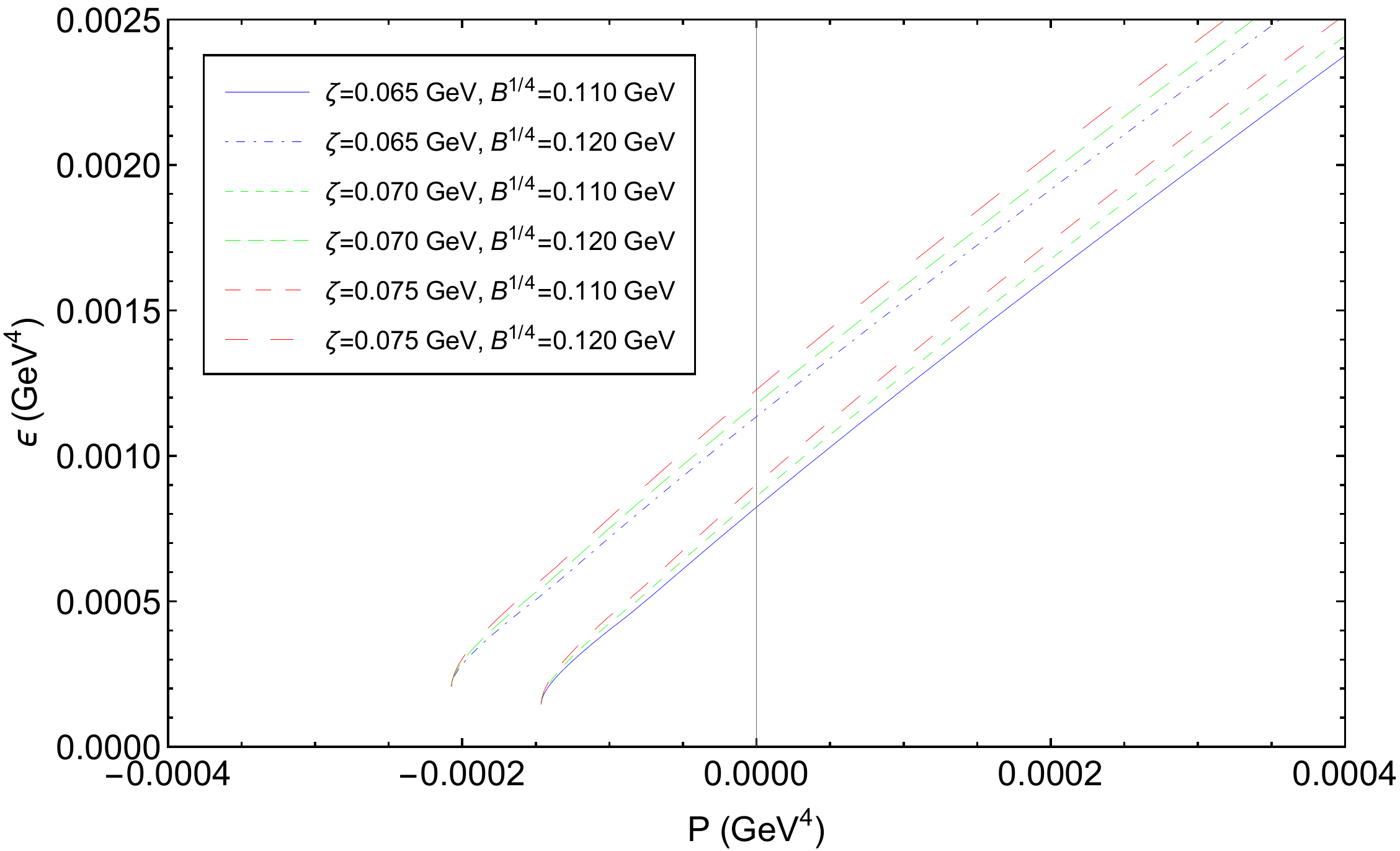}
	\caption{The relations between the energy density $ \varepsilon $ and the pressure density $ P $ for different $ \zeta $ and $ B $. }
	\label{fig6}
\end{figure}

The EOS of quark matter at zero temperature can be derived from statistical mechanics and reads \cite{41,42}
\begin{equation}\label{Eq.12}
P(\mu)=P(\mu)|_{\mu=0}+\int_{0}^{\mu}\mathrm{d}\mu'\rho(\mu'),
\end{equation}
where $ P(\mu)|_{\mu=0} $ is a negative term. The term $ P(\mu)|_{\mu=0} $ is so-called vacuum pressure which is related to the confinement property of QCD. However, due to the lack of comprehensive  understanding of QCD interactions, we can neither obtain this term from the first principles nor figure out how it is generated. Therefore, analogous to what researchers done in MIT bag model, we treat the vacuum pressure term as a phenomenological parameter and rewrite it as $ P(\mu)|_{\mu=0}=-B(B>0) $ in this paper. Then, the energy density can be deduced from the thermodynamic relation
\begin{equation}\label{Eq.13}
\varepsilon=-P+\sum_{i}\mu_{i}\rho_{i}.
\end{equation}
Combing  \eqref{Eq.12} and \eqref{Eq.13} with the discussion taken before, some significant relations can be obtained, as illustrated in figures \ref{fig4}, \ref{fig5} and \ref{fig6}, respectively.

The influence of parameters $ \zeta $,  parameter $B$ and baryon chemical potential $ \mu_{B} $ on the EOS of SQM are presented in figure \ref{fig4} and \ref{fig5}.  We can see that the pressure density $ P(\mu_{B}) $ and the energy density $ \varepsilon(\mu_{B}) $ have similar behaviors for baryon chemical potential $ \mu_{B} $: i. e.,  $ \mu_{B}<\mu_{Bc} $, the $ P(\mu_{B}) $ and $ \varepsilon(\mu_{B}) $ keep constants of $-B$ and $B$ respectively, and the reason for this phenomenon is the disappearance of quarks in this region; once $ \mu_{B}>\mu_{Bc} $, they become monotonically increasing functions of baryon chemical potential $ \mu_{B} $.  We exhibit the relation between the pressure density and energy density in figure \ref{fig6}. It can be found that the EOS with a larger $ \zeta $ and a larger $B$ becomes softer. In addition, there is a starting point of EOS, $(-B,B)$  in the plane of energy-pressure density, which is the   consequence of   \eqref{Eq.12}.

\begin{figure}[htbp]
	\centering
	\includegraphics[width=0.85\textwidth]{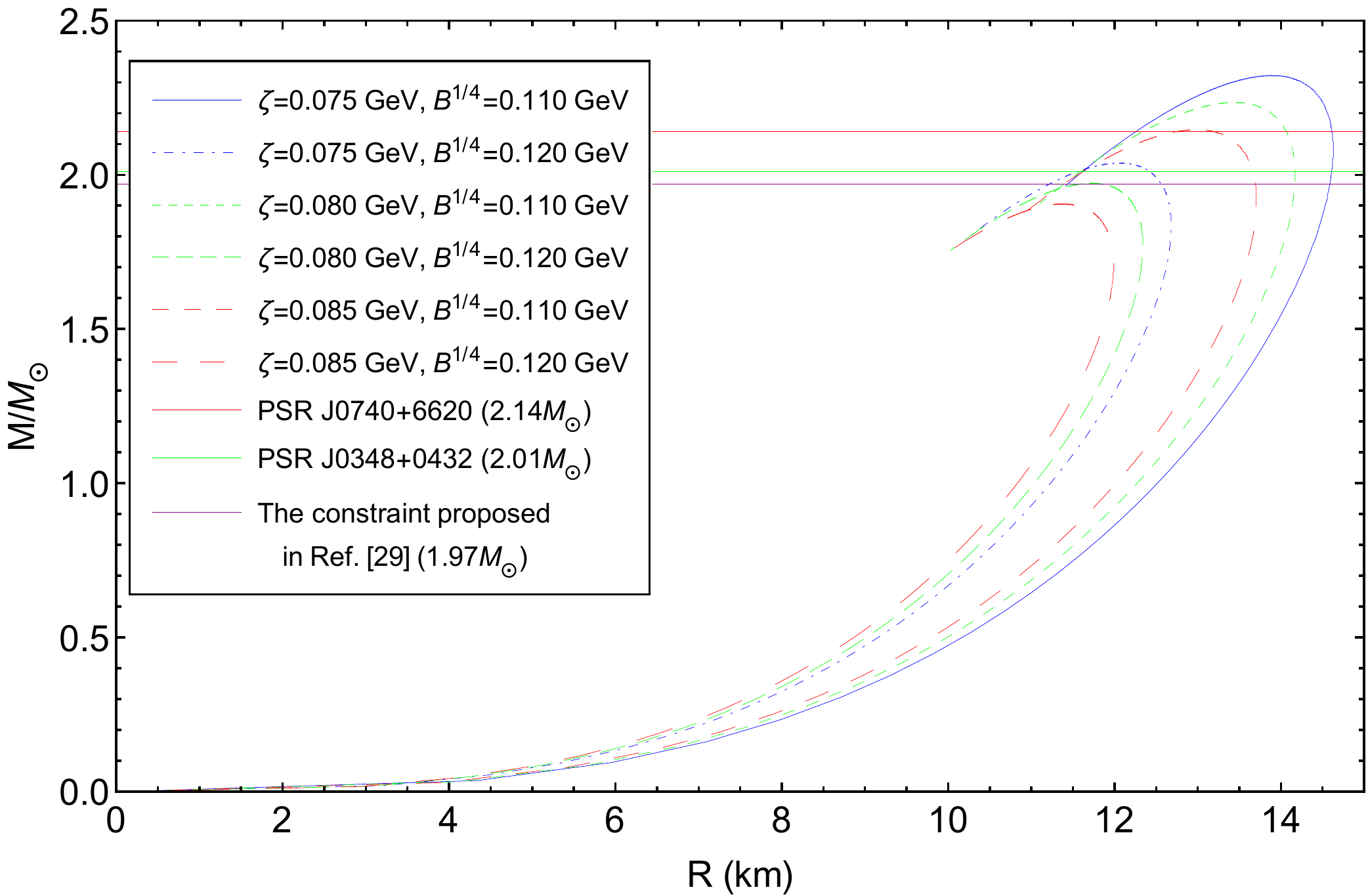}
	\caption{The mass-radius relations of strange quark stars for different $ \zeta $ and $ B $ with the constraints required by PSR J$0740+6620$, PSR J$0348+0432$.}
	\label{fig7}
\end{figure}

Utilizing the EOS of SQM obtained with quasi-particle model, we can get the mass-radius relations (in units of $G=c=1$) by integrating the TOV equations \cite{43,44,45}
\begin{equation}\label{Eq.14}
\frac{\mathrm{d}P(r)}{\mathrm{d}r}=-\frac{(\varepsilon+P)(M+4\pi r^{3}P)}{r(r-2M)} 
\end{equation}
and 
\begin{equation}\label{Eq.15}
\frac{\mathrm{d}M(r)}{\mathrm{d}r}=4\pi r^{2}\varepsilon,
\end{equation}
and the result is exhibited in figure \ref{fig7}. As a comparison, we also show the maximum mass constraints required by PSR J$0740+6620$ $(M=2.14^{+0.10}_{-0.09}~M_{\odot})$ and PSR J$0348+0432$ $(M=2.01\pm0.04~M_{\odot})$, as well as the restriction proposed in   \cite{28} in this figure. It is easy to find in figure \ref{fig7} that a larger maximum mass will be generated with a stiffer EOS. Furthermore, we can find that the EOSs of quasi-particle have the ability to yield a maximum mass larger than the latest observation results. The maximum masses and the corresponding radii of strange quark stars with different $ \zeta $ and $ B $ are calculated, and   the results are exhibited in table \ref{table1}. From this table we can see clearly that the radius of the strange quark star with a mass of near $ 2 $ $ M_{\odot} $ closes to $ 12 $ km. The results are in agreement with the discussions in   \cite{46,47,48,49}.

\begin{table}[htbp]
	\centering
	\caption{The radii $R$ and the maximum masses $ M_{max} $ of strange quark stars with different $ \zeta $ and $ B $.}
	\label{table1}
	\begin{tabular}{cccc}
		\hline\hline
	$\zeta$(GeV)&$ B^{1/4} $(GeV)&$ R $(km)&$ M_{max}(M_{\odot}) $\\
	\hline
	0.078&0.120&11.84&2.00\\
	0.079&0.119&11.95&2.01\\
	0.080&0.118&12.05&2.02\\
	0.081&0.117&12.16&2.03\\
	0.082&0.116&12.20&2.04\\
	\hline\hline
   \end{tabular}
\end{table}

According to the General Relativity, the gravitational fields of objects are not uniform or constant but rely on the motion state and position of gravitating bodies, which is so-called "gravitational effects" . Hence, for a gravitational field, there is a deviation from uniformity at nearby points caused by gravitational effects, and the concept of "tidal gravity" is used to refer to such a deviation. Therefore, with regard to a binary stars system, each star will be deformed by the tidal field of its companion because of the presence of tidal gravity. This phenomenon can be described by the physical quantity "tidal deformability", which is defined as the ratio between one star's induced mass quadrupole moment and the tidal field of its companion \cite{45}.

In order to calculate the tidal deformability $ \varLambda $ of a quark star in the case of low-spin prior, we   adopt the relation of
\begin{equation}\label{Eq.16}
k_{2}=\frac{3}{2}\varLambda C^{5},
\end{equation}
where $ C=M/R $ represents the compactness of a star. Here, the physical quantity $ k_{2} $ is the dimensionless tidal Love number for $l=2$ which describes how difficultly a star can be deformed by an external tidal field, and it can be expressed as
\begin{equation}\label{Eq.17}
  \begin{aligned}
  k_{2}&=\frac{8C^{2}}{5}(1-2C)^{2}[2+2C(y-1)-y]\\
  &\times\lbrace 2C[6-3y+3C(5y-8)]\\
  	&+4C^{3}[13-11y+C(3y-2)+2C^{2}(1+y)]\\
  	&+3(1-2C)^{2}[2-y+2C(y-1)]\ln(1-2C)\rbrace ^{-1}.
  \end{aligned}
\end{equation}
In   \eqref{Eq.17}, the symbol $y$ is defined as
\begin{equation}\label{Eq.18}
y=\frac{R\beta(R)}{H(R)}-\frac{4\pi R^{3}\varepsilon_{0}}{M}
\end{equation}
and related to metric function $ H(r) $ and surface energy density $ \varepsilon_{0} $. Due to the introduction of negative vacuum pressure, there is a nonzero energy density just around the surface of the star as we can see in figure \ref{fig6}. To derive the metric function, we are supposed to integrate the following differential equations
\begin{eqnarray}\label{Eq.19}
\frac{\mathrm{d}H(r)}{\mathrm{d}r}=\beta,\\
\frac{\mathrm{d} \beta(r)}{\mathrm{d} r}=& 2\left(1-2 \frac{M}{r}\right)^{-1} H\Bigg\{-2 \pi[5 \epsilon+9 P+f(\epsilon+P)] \nonumber\\
&\left.+\frac{3}{r^{2}}+2\left(1-2 \frac{M}{r}\right)^{-1}\left(\frac{M}{r^{2}}+4 \pi r P\right)^{2}\right\} \nonumber\\
&+\frac{2 \beta}{r}\left(1-2 \frac{M}{r}\right)^{-1}\left\{-1+\frac{M}{r}+2 \pi r^{2}(\epsilon-P)\right\},
\end{eqnarray}
where
\begin{equation}\label{Eq.21}
f=\frac{\mathrm{d}\varepsilon}{\mathrm{d}p}.
\end{equation}
The integration  will start from the center with the expansions $ H(r)=a_{0}r^{2} $ and $ \beta(r)=2a_{0}r $ as the radius $ r\to0 $. Since what we are concerned about is the ratio of $ \beta/H $, we can ignore the value of coefficient $ a_{0} $ and treat it as $ 1 $ for simplicity.

\begin{figure}[htbh]
	\centering
	\subfigure[]
	{
		\label{fig8a}
		\includegraphics[width=0.85\textwidth]{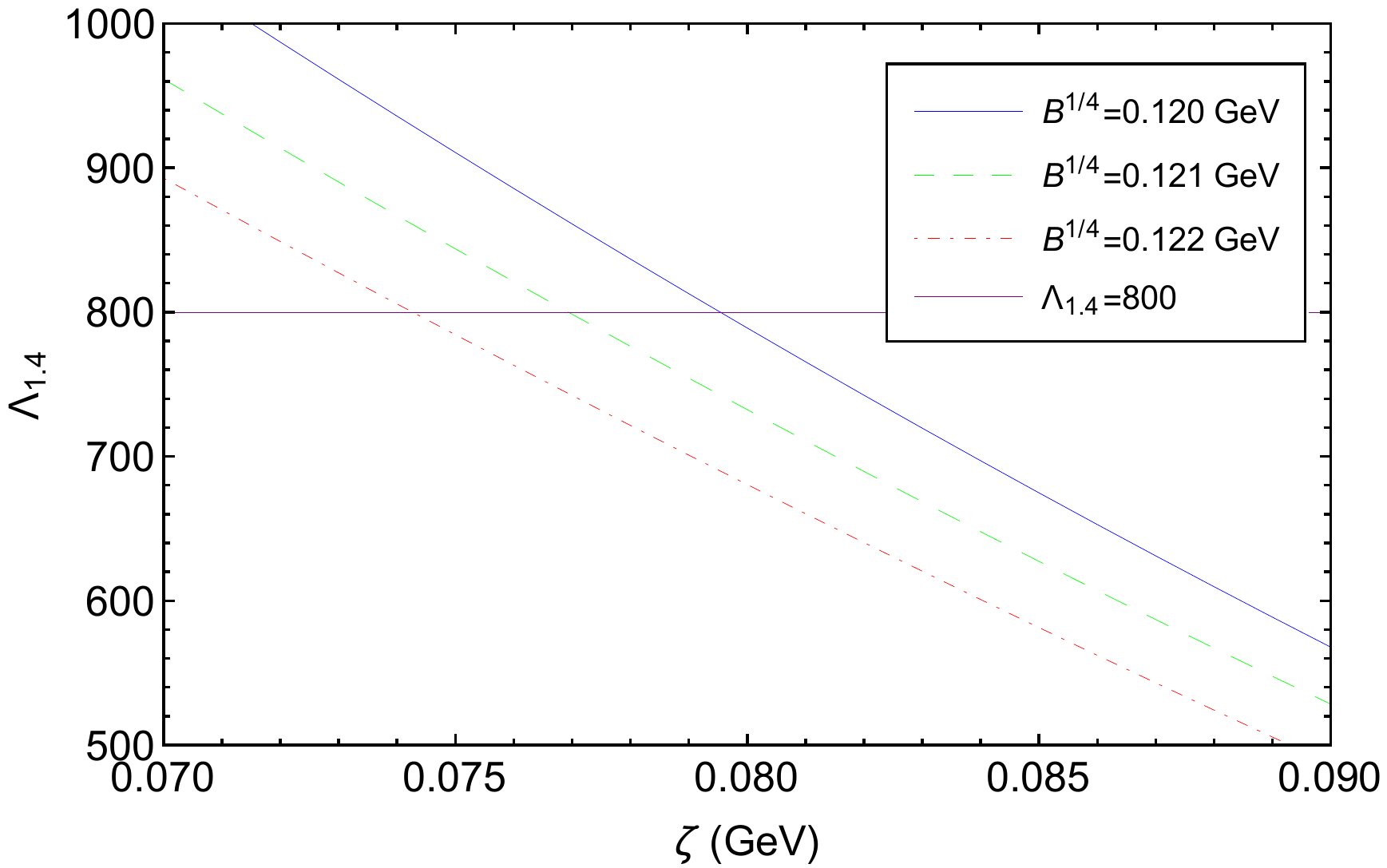}
}
     \subfigure[]
{
	\label{fig8b}
	\includegraphics[width=0.85\textwidth]{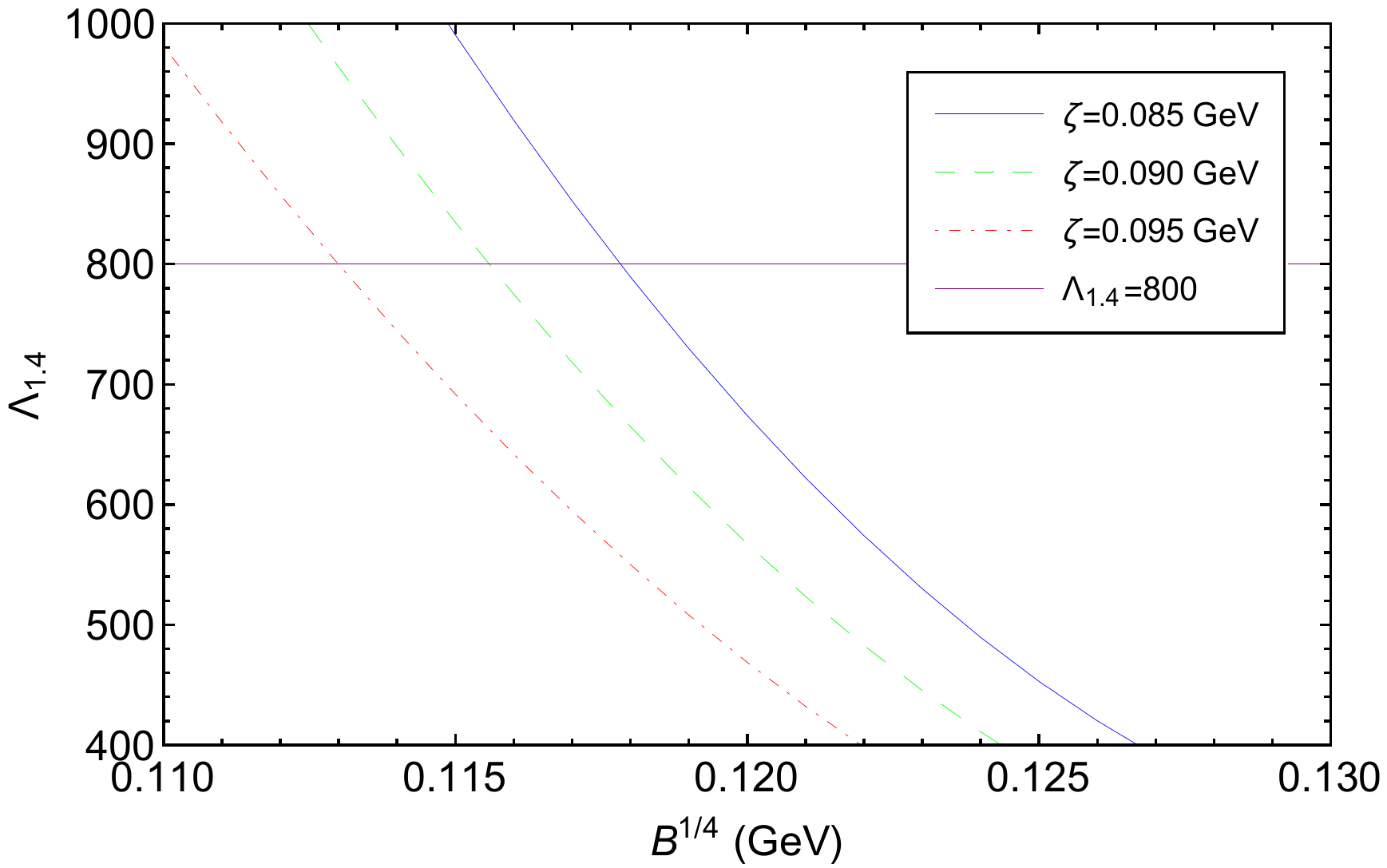}
}
   \caption{The relations between the $ \varLambda_{1.4} $ and parameter $ \zeta $ \subref{fig8a} as well as the relations between the $ \varLambda_{1.4} $ and parameter $ B $ \subref{fig8b}.}
   \label{fig8}
\end{figure}

Combining the discussion taken above with the EOS of quasi-particle model, the properties of strange quark stars with $ 1.4 $ $ M_{\odot} $ for different $ \zeta $ and $ B $ can be obtained, as shown in table \ref{table2}. From this table we can see that the strange quark star is less compact with a smaller $ \zeta $ and a smaller $ B $ for a given mass. This result means such a star is easier to be deformed by an external tidal field.   We illustrate  in figure \ref{fig8} the relations between the tidal deformability for a $ 1.4 $  $ M_{\odot} $ strange quark stars $ (\varLambda_{1.4}) $ and parameters $ \zeta $ as well as the relations between $ \varLambda_{1.4} $ and parameter $ B $. For comparison, we also show the constraint required by the  GW$ 170817 $ that the $ \varLambda_{1.4} $ should be smaller than $ 800 $ in the same figure. There is no doubt that a larger $ \zeta $ and a larger $ B $ corresponds to a smaller $ \varLambda_{1.4} $. Besides, we can find in Fig. \ref{fig8} that not all the parameter settings satisfy the constraint from gravitational wave observations.

\begin{table}[htbp]
	\centering
	\caption{The properties of strange quark stars with a mass of 1.4 $ M_{\odot} $ for different $ \zeta $ and $ B $, including the compactness $ C=M/R $, the Love number $ k_{2} $ as well as the tidal deformability $ \varLambda_{1.4} $.}
	\label{table2}
	\begin{tabular}{ccccc}
		\hline\hline
		$ \zeta $(GeV)&$ B^{1/4} $(GeV)&$ C=M/R $&$ k_{2} $&$ \varLambda_{1.4} $\\
		\hline
		0.120&0.078&0.171&0.187&833.682\\
		0.120&0.079&0.172&0.186&819.215\\
		0.120&0.080&0.173&0.184&787.925\\
		0.121&0.081&0.175&0.181&731.478\\
		0.121&0.082&0.177&0.177&678.982\\
		\hline\hline
	\end{tabular}
\end{table}

\begin{figure}[htbh]
	\centering
	\subfigure[]
	{
		\label{fig9a}
		\includegraphics[width=0.85\textwidth]{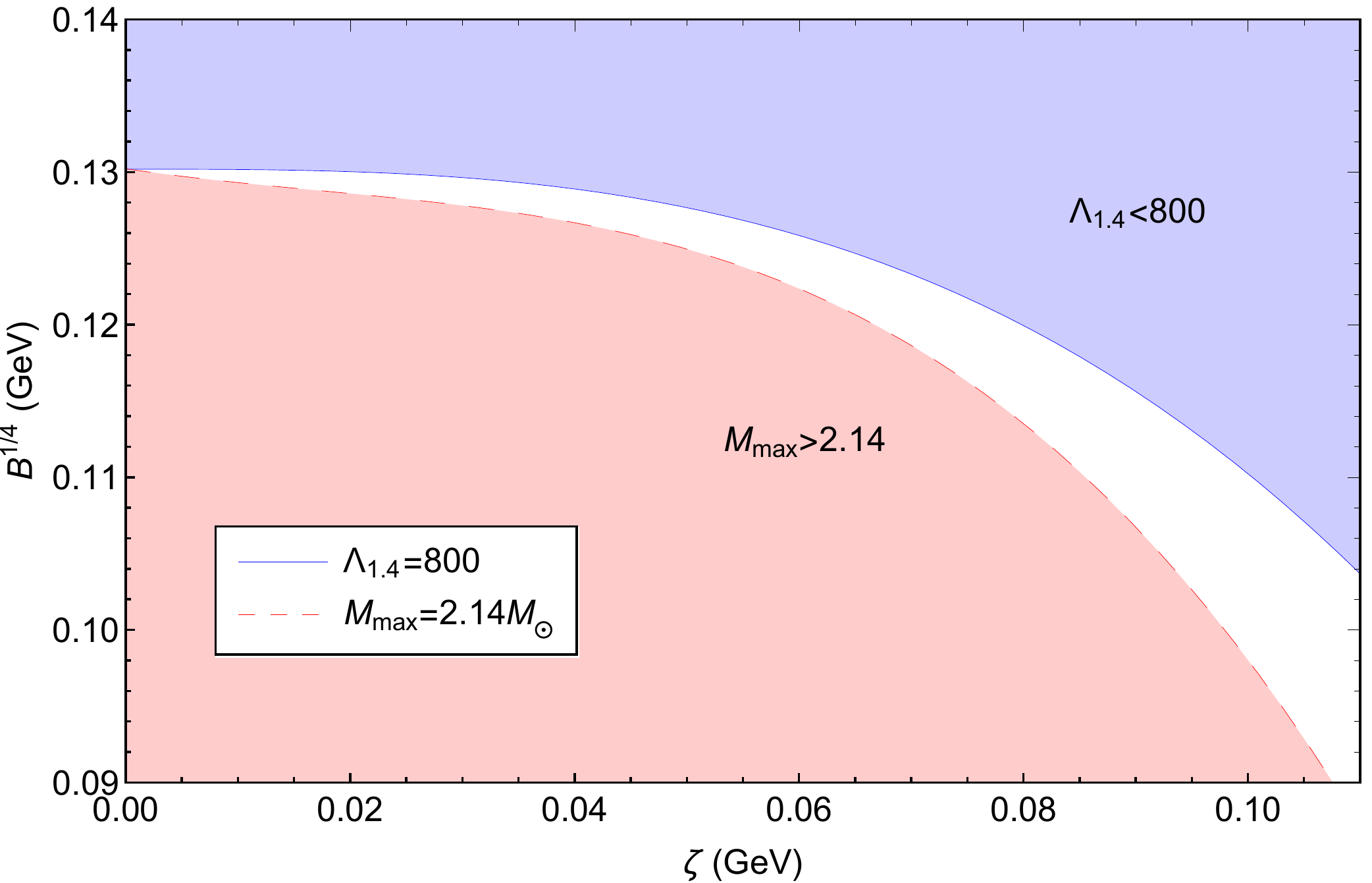}
	}
	\subfigure[]
	{
		\label{fig9b}
		\includegraphics[width=0.85\textwidth]{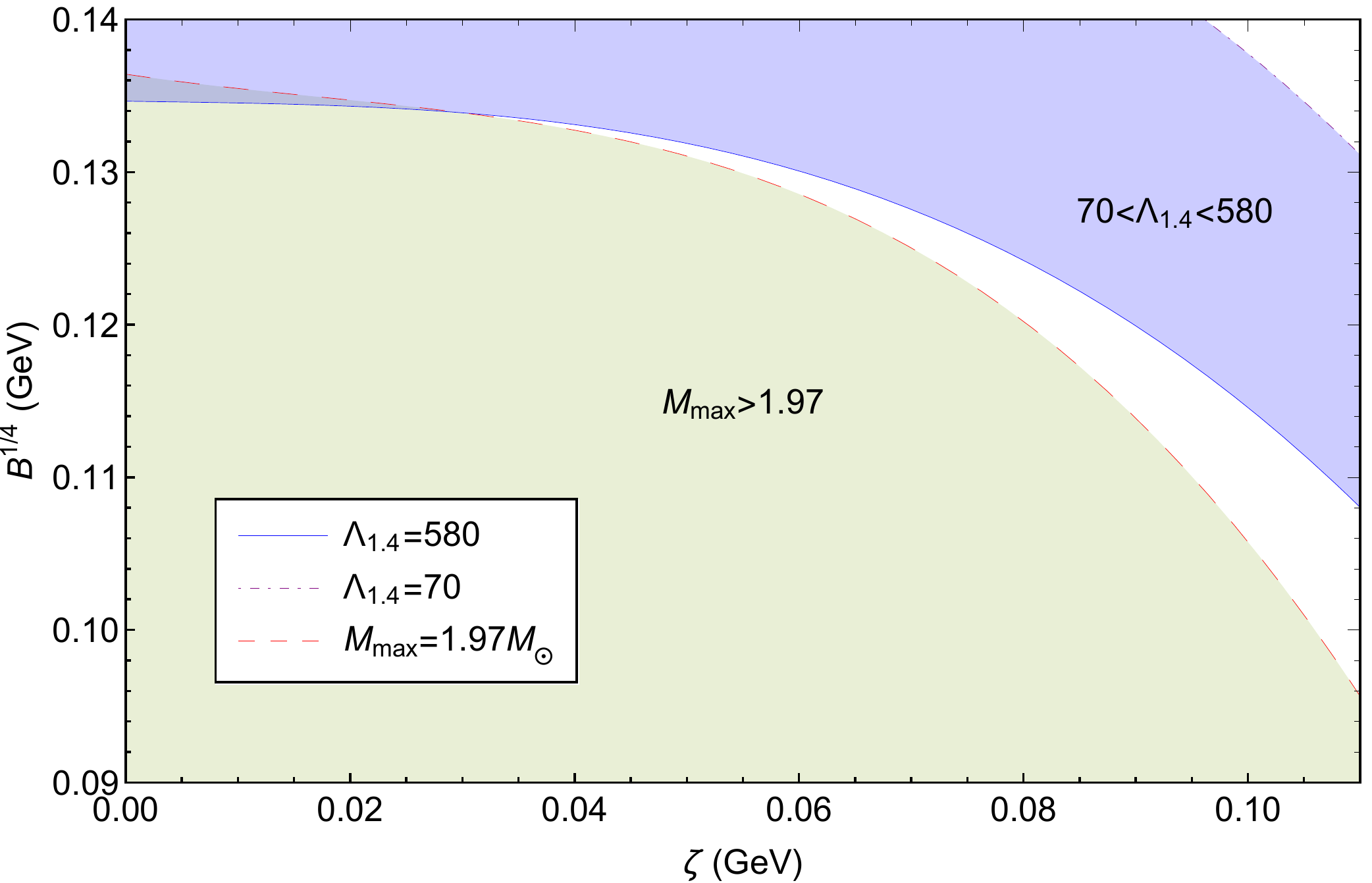}
	}
     \caption{The parameter space of the quasi-particle model based on observation results of PSR J$ 0740+6620 $ $ (M=2.14^{+0.10}_{-0.09}~M_{\odot}) $ and GW$ 170817 $ \subref{fig9a} and based on constraints proposed in   \cite{28} \subref{fig9b}. }
     \label{fig9}
\end{figure}

In figure \ref{fig9}, we illustrate the parameter space of the quasi-particle model basing on the astronomical observations and theoretical results respectively. In panel \subref{fig9a}, we constrain the parameters in view of the PSR J$ 0740+6620 $ $ (M=2.14^{+0.10}_{-0.09}~M_{\odot}) $ and GW$ 170817 $, and find that there is no coincident region for the feasible area of PSR J$ 0740+6620 $ and the feasible area of GW$ 170817 $. Namely, the constraint of GW$ 170817 $, which requires $ \varLambda_{1.4} $ is smaller than $ 800 $, and the constraint of PSR J$ 0740+6620 $$ (M=2.14^{+0.10}_{-0.09}~M_{\odot}) $, which requires a maximum mass above 2.14 $ M_{\odot} $, can not be met simultaneously. Similar to panel \subref{fig9a}, we also draw another parameter space in panel \subref{fig9b}, which is based on the requirement proposed in Ref. \cite{28}. In Ref. \cite{28}, researchers claimed that the EOS must ensure the tidal deformability $ \varLambda_{1.4}=190^{+390}_{-120} $ and support a maximum mass above 1.97 $ M_{\odot} $. It is clear to see from \subref{fig9b} that the two feasible areas do not coincide when the $ \zeta $ is larger than about $ 0.03 $ GeV. And in the region of $ \zeta<0.03 $ GeV, there is a small coincident area for the two constraints, which almost vanishes. However, according to Ref. \cite{50}, the quasi-particle model we use will work better in the large $ \zeta $ region in order to fit the LGT data; as for the area of small $ \zeta $, including the region below $ 0.03 $ GeV, the results may be unreliable to some extent, let alone the coincident region is very small compared to the whole parameter space.

It shows in figure \ref{fig9} that the parameter space of quasi-particle model can not meet the the constraints of PSR J$ 0740+6620 $$ (M=2.14^{+0.10}_{-0.09}~M_{\odot}) $ and GW$ 170817 $ simultaneously. Meanwhile, the parameter space also can not restrict the requirement proposed in   \cite{28} in a reliable region. Up to now, the increasing evidences indicate that the compact stars of large masses may be strange quark stars \cite{51,52,53,54}. On the basis of the results in figure \ref{fig9}, we think it is probably improper to regard the compact star with a mass of $ 1.4 $ $ M_{\odot} $ as strange quark star. Naturally, we assume that the properties of compact stars with small mass (including $ 1.4 $ $ M_{\odot} $) look more like neutron stars formed by hadronic matter rather than strange quark stars. This hypothesis is consistent with the conclusion of   \cite{28}, in which the researchers claimed that they found the quark matter exist within the compact stars with $ 2 $ $ M_{\odot} $ but not present within the compact stars with $ 1.4~M_{\odot} $.

\section{Summary}
\label{sum}
In this paper, we have studied the properties of the strange quark stars with quasi-particle model basing on the latest astronomical observations and theoretical results. Taking the chemical equilibrium of weak interaction and charge neutrality into account, we have got the baryon number density and the variation of constituents for different $ \zeta $. We have found that the proportions of quarks are irrelevant with $ \zeta $. With the introduction of parameter $ B $ for the vacuum pressure, we have established the EOS of SQM.   With proper choices of $B$ and $\zeta$, it is able to generate a quark star with the maximum mass larger than $ 2.14~M_{\odot} $, which matches the results of PSR J$ 0740+6620 $ $ (M=2.14^{+0.10}_{-0.09}~M_{\odot}) $ and PSR J$ 0348+0432 $ $ (M=2.01\pm0.04~M_{\odot}) $. We have calculated the properties of strange quark star with $ 1.4~M_{\odot} $, including the compactness $ C $, the Love number $ k_{2} $ and the tidal deformability $ \varLambda_{1.4} $. It is found that   a softer EOS corresponds to a more compact quark star, which is more difficult to be   deformed. 

Finally, we have illustrated the parameter space of the quasi-particle model basing on the astronomical observations and theoretical suggestions respectively.
It   is   found that the parameter space can not meet the constraint of GW$ 170817 $  which requires $ \varLambda_{1.4} $ is smaller than $ 800 $, and the constraint of PSR J$ 0740+6620 $ $ (M=2.14^{+0.10}_{-0.09}~M_{\odot}) $  which require a maximum mass of at least $ 2.14~M_{\odot} $, simultaneously. Similarly, the model also can not satisfy the constraints proposed in  \cite{28}, in which the researchers claimed  that the EOS must ensure the tidal deformability $ \varLambda_{1.4}=190^{+390}_{-120} $ and support a maximum mass above $ 1.97~M_{\odot} $. Since more and more evidences indicate that the compact stars and pulsars with large masses are   quark stars, we assume that the properties of compact stars and pulsars with a small mass (including $ 1.4~M_{\odot}  $) looks more like a neutron star formed by hadronic matter rather than a strange quark star basing on the analysis of figure \ref{fig9}. 
A recent work assuming the strange quark matter is in the color-flavor locked phase has found that the color superconductivity gap is poorly constrained by those observed global   properties of a $1.4M_{\odot}$ star \cite{angli}.
 As such,  it is suggested to use   hadronic EOS   in exploring the properties of low-mass compact stars  while to use  pure quark EOS or hybrid EOS  in studying the compact stars with larger mass.


\section*{References}


\begin{thebibliography}{200}
\bibitem{1}   Buballa M 2005 \textit{Phys. Rep.} \textbf{407}  205  
\bibitem{2}   Luo  X  F  and  Xu N   2017 \textit{Nucl. Sci. Tech.} \textbf{28}  112  
\bibitem{3}  Ivanenko D and   Kurdgelaidze D  F  1969 \textit{Lett. Nuovo Cim.} \textbf{2}  13  
\bibitem{4}  Itoh N  1970 \textit{Prog. Theor. Phys.} \textbf{44}, 291 
\bibitem{5}   Iwamoto N 1980 \textit{Phys. Rev. Lett.} \textbf{44}, 1637  

\bibitem{6}   Bodmer A  R   1971 \textit{Phys. Rev.} D \textbf{4}, 1601  
\bibitem{7}    Haensel P,   Zdunik J L  and  Schaeffer R 1986 \textit{Astron.Astrophys.} \textbf{160}, 121  
\bibitem{8}  Alcock C,   Farhi E  and   Olinto A 1986 \textit{Astrophys.} J. \textbf{310}, 261  
\bibitem{Terazawa}   Terazawa  H 1989 \textit{J. Phys. Soc. Jpn.} \textbf{58}, 3555  
\bibitem{24}  Witten E 1984 \textit{Phys. Rev.} D \textbf{30}, 272 
\bibitem{55}   Kuerban A,   Geng J J,   Huang Y F,   Zong  H S and   Gong H 2019 Close-in Exoplanets as Candidates of Strange Quark Matter Objects arXiv:1908.11191 [astro-ph.HE]  
\bibitem{9}   Ozel F 2006 \textit{Nature} \textbf{441}, 1115  
\bibitem{10}  Alford M et al 2007 \textit{ Nature} \textbf{445}, E7  

\bibitem{11}   Chodos A,    Jaffe R  L, Johnson K,   Thorn C  B and   Weisskopf V  F  1974 \textit{Phys. Rev.} D \textbf{9}, 3471  
\bibitem{12}   Alford M,  Braby M,   Paris M  and   Reddy S 2005 \textit{Astrophys.} J. \textbf{629}, 969  
\bibitem{13}  Alcock C,   Farhi E, and   Olinto A 1986 \textit{Astrophys.} J. \textbf{310} 261  
\bibitem{14}   Zhou E P,  Zhou X, and   Li A 2018 \textit{Phys. Rev.} D \textbf{97}, 083015  
\bibitem{15}  Wang Q W,   Shi C, and  Zong H S   2019 \textit{Phys. Rev.} D \textbf{100}, 123003  

\bibitem{16}   Wang Q W,  Xia Y and   Zong H S 2018 \textit{Mod. Phys. Lett.}  A \textbf{33}, 1850232 (2018).
\bibitem{17}  Fan Z Y,   Fan W K,   Wang Q W and  Zong H S  2017 \textit{Mod. Phys. Lett.} A \textbf{32}, 1750107  
\bibitem{19}  Menezes D P,  Providncia C, and   Melrose D  B  2006 \textit{J. Phys.} G \textbf{32}, 1081  
\bibitem{20}  Peshier A,   Kampfer B,  Soff G 2000 \textit{Phys. Rev.} C \textbf{61}, 045203  

\bibitem{21}   Szabo K K,   Toth A  I 2003 \textit{JHEP} \textbf{06}, 008  
\bibitem{22}  Plumari S,   Alberico W M,   Greco V and   Ratti C 2011 \textit{Phys. Rev.} D \textbf{84}, 094004 
\bibitem{23}  Ma H H,   Dudek D M,   Lin K  and et al 2018 A quasi-particle model with a phenomenological critical point arXiv:1804.06797 [nucl-th]  
\bibitem{25}  Cromartie H  T,  Fonseca E,   Ransom S  M and et al 2019 Nat. Astron. \textbf{4}, 72  
\bibitem{26}  Antoniadis J et al 2013 \textit{Science} \textbf{340}, 1233232  
 \bibitem{27}   Abbott B  P  and et al 2017
 \bibitem{28}  Annala E,   Gorda T,   Kurkela A,   Nattila J and   Vuorinen  A 2020 Nature Phys. doi:10.1038/s41567-020-0914-9  \textit{Phys. Rev. Lett.} \textbf{119}, 161101  
\bibitem{29}   Fattoyev F  J,   Horowitz C J,   Piekarewicz J and   Reed B  2020 GW190814: Impact of a 2.6 solar mass neutron star on nucleonic equations of state  arXiv:2007.03799 [nucl-th] 
\bibitem{30}    Tan H,   Noronha-Hostler J  and   Yunes N  2020 Neutron Star Equation of State in light of GW190814 arXiv:2006.16296 [astro-ph.HE]  

\bibitem{31}  Bannur V  M  2007 \textit{Eur. Phys. J.} C \textbf{50}, 629  
\bibitem{32}  Bannur V  M  2007 \textit{Phys. Lett. } B \textbf{647}, 271  
\bibitem{33}  Bannur V  M  2007 \textit{Phys. Rev.  } C \textbf{75}, 044905  
\bibitem{34}  Bannur V  M  2007 \textit{JHEP} \textbf{09}, 046  
\bibitem{35}  Bannur V  M  2008 \textit{Phys. Rev.} C \textbf{78}, 045206  

\bibitem{36}   Ma H H and  Qian W L  2018  \textit{Braz. J. Phys.} \textbf{48}, 160  
\bibitem{59}  Rebhan A and Romatschke P  2003  \textit{Phys. Rev.} D \textbf{68}, 025022  
\bibitem{60}  Schneider R A 2003 The QCD running coupling at finite temperature and density    arXiv:hep-ph/0303104 [hep-ph]  
\bibitem{37} Halasz M A,   Jackson A D,   Shrock R E,   Stephanov M A and  Verbaarschot  J M 1998 \textit{Phys. Rev.} D \textbf{58}, 096007  
\bibitem{38} Tian Y L, Yan Y,   Li H,   Luo X L and   Zong H S 2012 \textit{Phys. Rev.} D \textbf{85}, 045009 
\bibitem{39}   Simji P 2020 \textit{Int. J. Mod. Phys.} A \textbf{35},  2050064  


\bibitem{41} He M,   Feng H  T,   Sun W M  and  Zong H S  2007 \textit{J. Phys. G} \textbf{34}, 2655 
\bibitem{42}   Zong  H S  and   Sun W M  2008 \textit{Phys. Rev.} D \textbf{78}, 054001  
\bibitem{43}  Hinderer T 2008 \textit{Astrophys.} J. \textbf{677}, 1216  
\bibitem{44}  Damour T and   Nagar A 2009 \textit{Phys. Rev.} D \textbf{80}, 084035  
\bibitem{45}   Yagi K and  Yunes N 2013  \textit{Science} \textbf{341}, 365  

\bibitem{46}  Li B L,   Cui Z F,   Yu Z H,  Yan Y,   An S  and   Zong H S  2019  \textit{Phys. Rev.} D \textbf{99}, 043001  
\bibitem{47}  Zong H S   and   Sun W M  2008  \textit{Phys. Rev.} D \textbf{78}, 054001  
\bibitem{48}   Annala E,   Gorda T, Kurkela A and   Vuorinen A 2018 \textit{Phys. Rev. Lett.} \textbf{120}, 172703  
\bibitem{49}   Wang Q W,  Shi C, Yan  Y  and  Zong H S   2019 Exploring hybrid equation of state with constraints from tidal deformability of GW170817  arXiv:1912.02312 [hep-ph]  
\bibitem{50}   Bannur V  M   2012 \textit{Int. J. Mod. Phys.} E \textbf{21}, 1250090  

\bibitem{51}  Li C M,   Zuo S Y,   Yan Y,  Zhao Y P,   Wang F,  Huang Y F and  Zong H S   2020 \textit{Phys. Rev.} D \textbf{101}, 063023  
\bibitem{52}  Jokela  N,  Jarvinen M  and  Remes J  2019 \textit{JHEP} \textbf{03}, 041  
\bibitem{53}   Pinkanjanarod S and  Burikham P  2020 Massive neutron stars with multiquark cores  arXiv:2007.10615 [nucl-th]   
\bibitem{54} Otto K,  Oertel M  and  Schaefer B J  2020  Nonperturbative quark matter equations of state with vector interactions arXiv:2007.07394 [hep-ph]  

\bibitem{angli}   Li A,  Jiang J L,  Tang S  P,  Miao  Z   Q,   Zhou E   P and  Xu R X  2020  Constraints from LIGO/Virgo and NICER on quark star equation of state  arXiv:2009.12571 [hep-ph]  

\end{thebibliography}
\end{document}